# SCAF – AN EFFECTIVE APPROACH TO CLASSIFY SUBSPACE CLUSTERING ALGORITHMS

Sunita Jahirabadkar[1] and Parag Kulkarni[2]

[1]Department of Computer Engineering & I.T., College of Engineering, Pune, India
`sunita.jahirabadkar@gmail.com`
[2]Department of Computer Engineering & I.T., College of Engineering, Pune, India
`parag.kulkarni@eklatresearch.in`

## ABSTRACT
*Subspace clustering discovers the clusters embedded in multiple, overlapping subspaces of high dimensional data. Many significant subspace clustering algorithms exist, each having different characteristics caused by the use of different techniques, assumptions, heuristics used etc. A comprehensive classification scheme is essential which will consider all such characteristics to divide subspace clustering approaches in various families. The algorithms belonging to same family will satisfy common characteristics. Such a categorization will help future developers to better understand the quality criteria to be used and similar algorithms to be used to compare results with their proposed clustering algorithms. In this paper, we first proposed the concept of SCAF (Subspace Clustering Algorithms' Family). Characteristics of SCAF will be based on the classes such as cluster orientation, overlap of dimensions etc. As an illustration, we further provided a comprehensive, systematic description and comparison of few significant algorithms belonging to "Axis parallel, overlapping, density based" SCAF.*

## KEY WORDS
*Axis parallel clusters, Density based clustering, High dimensional data, Subspace clustering,*

## 1. INTRODUCTION

Clustering is the most common data mining process which aims at dividing datasets into subsets or clusters in such a way that the objects in one subset are similar to each other with respect to a given similarity measure, while objects in different subsets are dissimilar [1]. Clustering is commonly and heavily used in variety of application areas such as in medical science, environmental science, astronomy, geology, business intelligence and so on [2]. It helps users in understanding natural grouping in a dataset or structure of the dataset. Clustering is also treated as a form of data compression. Thus, in general, clustering can be treated as a first step in various data processing methods such as classification, indexing, data compression, etc.

While lot of work has been already done in the area of clustering [1, 3, 4], new approaches need to be proposed to cope with the modern capabilities of huge data generation. Clustering the real world datasets need to deal with objects modeled by high dimensional data, where each object is described by hundreds or thousands of attributes. For instance, there are many computer vision applications, such as motion segmentation, face clustering with varying illumination, pattern classification, temporal video segmentation etc. In such applications, image data is very high dimensional. Another example of high dimensional data can be found in the area of molecular biology [5] and CAD (Computer Aided Design) databases. However, such high dimensional data initiates different challenges for conventional clustering approaches [6]. In particular, the

DOI : 10.5121/ijdkp.2013.3205                                                              69

International Journal of Data Mining & Knowledge Management Process (IJDKP) Vol.3, No.2, March 2013

traditional clustering algorithms [7, 8] fail in such cases due to the inherent sparsity of the data objects and do not produce meaningful clusters.

In high dimensional data, clusters are embedded in various subsets of the entire dimension space [9]. A new research area of high dimensional data clustering detects such clusters embedded in different, variable length subspaces. There exist lot of approaches for subspace clustering and numerous algorithms are being proposed nearly every day. However, as this field is yet latest and emerging, there is no common ground on the basis of which we can bring all the algorithms on a universal platform, so as to compare their results. Hence, we need to effectively classify all these approaches to enable compare their various characteristics / features.

Few surveys of high dimensional data clustering approaches are available in literature [5, 10, 11, 12, 13]. An extremely comprehensive survey [10] illustrates different terminologies used and discusses various assumptions, heuristics or intuitions forming the basis of different high dimensional data clustering approaches. However, there is a need of classifying all subspace clustering approaches using multiple parameters which will group algorithms of similar characteristics in one family. Thus, the purpose of this paper is to provide the concept of clustering family SCAF (Subspace Clustering Algorithms' Family). For example, "Axis parallel, overlapping, bottom up, density based subspace clustering algorithms" will form one family. A researcher, who is using the same techniques to develop his / her clustering algorithm, will compare only with the algorithms belonging to this family.

This paper is structured as follows. The Remaining part of this section provides a short lead up to challenges involved in high dimensional data clustering and traditional methods of dimensionality reduction. Section 2 presents a detailed survey of various existing classification schemes of subspace clustering approaches followed by the introduction to the notion of SCAF. For ready reference, section 3 presents a comparative study of few significant algorithms belonging to "Axis parallel, overlapping, density based" SCAF. A comparative chart is shown indicating working principles, heuristics used, shape and size of the clusters, run time, accuracy as well as limitations etc. of different algorithms, followed by the conclusion in section 4.

## 1.1 Challenges of High Dimensional Data Clustering

There are three main challenges to high dimensional data clustering.

### 1.1.1 Curse of Dimensionality or Sparse Data

Many of the research streams like statistics, machine learning, data mining etc. contribute largely in the vigorous development of data clustering. However, these streams focus mainly on distance-based cluster analysis. Intuitively, the distance indicates similarity or dissimilarity between two data points (Figure 1). The distance between any two data objects is usually measured by a distance metric using the differences between the values of the attributes [2]. Typically, few clustering approaches measure Euclidian distance, using one of the $L_p$-norms. Others use a pattern, based on the behavior of attribute values to decide the similarity between data points. As such, different approaches result in different clustering models [10].

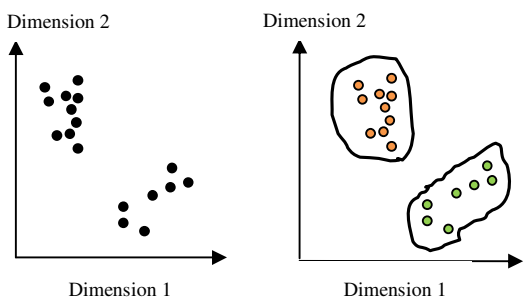

Figure 1. Dividing data in 2-clear groups using the distance between data points

Determination of simple Euclidian distance may not be useful in clustering high dimensional data [6]. Traditional clustering algorithms consider all the dimensions of an input dataset to measure such a distance between any two data points. While dealing with high dimensional data, clustering faces the hitches of 'Curse of dimensionality' [15] and the related sparsity problems. These hitches result from the fact that a fixed number of data points become increasingly 'sparse' as dimensionality increases.

In effect, the amount of data to sustain a given density increases exponentially with the dimensionality of the input space. On the other hand, the sparsity increases exponentially given a constant amount of data, with data points tending to become equidistant from one another. This will badly affect any clustering method which is based on either density or the distance between data points. Curse of dimensionality is illustrated in a simple way in Figure 2.

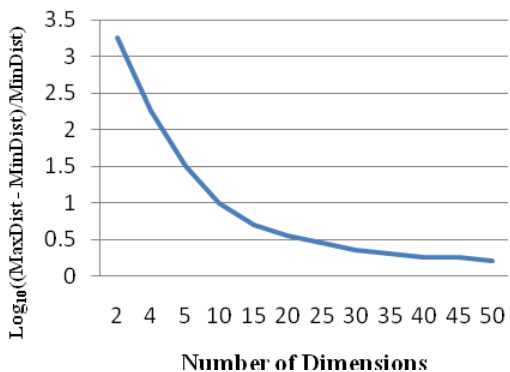

Figure 2. The curse of dimensionality

In Figure 2, randomly generated 200 data points have been used and difference between maximum distance and minimum distance among every pair of points is computed. Ref. [6, 16] show that, for certain data distributions, the relative difference of the distances between closest and farthest data points tends to 0 (for $L_d$ metric, and $d \geq 3$) as number of dimensions increases. i.e.

$$\lim_{d \to \infty} (\frac{MaxDist - MinDist}{MinDist}) = 0 \qquad (1)$$

where d is the number of dimensions [17].

Thus, eq. (1) shows the potential problems in high dimensional data clustering, in the cases where the data distribution generates relatively uniform distance between data points.

**1.1.2 Irrelevant Dimensions**

Apart from the curse of dimensionality, high dimensional data contains many of the dimensions often irrelevant to clustering or data processing [9]. These irrelevant dimensions confuse clustering algorithms by hiding clusters in noisy data. In such a case, the common approach is to reduce the dimensionality of the data, of course, without losing important information. Thus, clustering is often preceded by a 'feature selection' step which attempts to remove irrelevant features from the data. However, in high dimensional data, clusters are embedded in various subspaces. One dimension may be useful in some combinations of subspace for clustering the data, whereas, it can be irrelevant in some other subspace formation. Thus a global filtering approach for feature selection is not feasible.

**1.1.3 Correlations among Dimensions**

As there is large number of dimensions, there is some correlation among attributes. So, it may be possible that the clusters are not aligned to axis parallel, but can be arbitrarily oriented.

These problems make the average density in the data space quite low. Not only that the density in data space is low, but the noise values are also uniformly distributed in high dimensional space [18]. Thus, it is not effective to search for clusters in high dimensional data using the traditional clustering approaches.

There can be two ways to deal with the problem of high dimensionality. The first way would be to use variety of techniques performing dimensionality reduction prior to clustering. It reduces the number of dimensions in the given data, so that one can use existing clustering approaches without changing the meaning of the data. The other way is known as subspace clustering which solves the problems of high dimensional data clustering by building clusters hidden in the lower dimensional subspaces of the original dimension space.

**1.2 Dimensionality Reduction**

Fundamental techniques to eliminate irrelevant dimensions can be considered as 'Feature Selection' or 'Feature Transformation' techniques [19].

Feature transformation methods project the higher dimensional data onto a smaller space while preserving the distance between the original data objects. These methods use aggregation, dimensionality reduction etc. for summarizing data and creating linear combinations of the attributes. Such techniques enhance the data analysis in some cases as they are effective in removing noise. The commonly used methods are Principal Component Analysis [3, 20], Singular Value Decomposition [16] etc. There is a major limitation to feature transformation approach. This approach does not actually remove any of the attributes and hence the information from not-so-useful dimensions is preserved, making the clusters less meaningful. Thus the method is best suited only when there are not irrelevant dimensions in the data.

Feature selection methods try to remove some of the irrelevant dimensions from the high dimensional feature space. These methods require searching through various attribute subsets and need evaluating these subsets against some given criteria for clustering. Few popular feature selection techniques are discussed in ref. [21, 22, 23, 24]. The problem with these techniques is that they convert many dimensions to a single set of dimensions which later makes it difficult to interpret the results. Also, these approaches are inappropriate if the clusters lie in different subspaces of the dimension space.

While, all these dimensionality reduction techniques are quite successful on a large set of database applications, they face difficulty when clusters exist in various subspaces of the dimension space.

### 1.3 Subspace Clustering

Subspace clustering is the commonly used extension to feature selection which searches for groups of clusters in different subsets of the entire feature space of a high dimensional dataset. Subspace clustering thus finds interesting sub-sets of dimensions in the entire feature space and then the clusters hidden in those subspaces. Formally, a subspace cluster is defined as (Subspace of the feature space, Subset of data points). Or

$$C = (O, S) \text{ where } O \subseteq DB, S \subseteq D$$

Here, C is a subspace cluster, O is a set of objects in given database DB and S is a subspace projection of the dimension space D.

Subspace clustering needs to face two challenges - First to search for the relevant subspaces and then to find out the clusters in each of these subspaces [10]. As the search space for subspaces in general is infinite, it is necessary to apply some heuristic approach to make the processing feasible [15, 16]. Heuristic approach decides the characteristics of the algorithm. Once the subspaces with higher probability of comprising good quality clusters are identified, any clustering algorithm can be applied to find the hidden clusters.

Motivated with this fact that the clusters can be discovered in various subspaces of the dimension space, a significant amount of research has been presented aiming to discover such clusters of variable length subspaces. The first subspace clustering algorithm is proposed by R. Agrawal [9]. Later, many significant algorithms have been presented [14, 27, 28, 29, 30, 31, 32]. While all these approaches organize data objects into groups, each of them uses different methodologies to define clusters. They make different assumptions for input parameters. They define clusters in dissimilar ways as overlapping / non-overlapping, fixed size and shape / varying size and shape and so on. The choice of a search technique, such as top down / bottom up, can also determine the characteristics of the clustering approach.

In addition to the expectations that the subspace clustering algorithms should be efficient and produce high quality, interpretable clusters; they must also be scalable with respect to the number of objects and number of dimensions. As such, it is nearly impossible to define a universal measure of quality to be used to compare the clustering results. Hence, a proper categorization of all such algorithms can help future developers to better understand the quality criteria to be used to test their proposed clustering algorithms.

The next section talks about various classification schemes of subspace clustering approaches.

## 2. CLASSIFICATION SCHEMES

Existing subspace clustering approaches can be categorized using various classification schemes. There exist few survey papers which classify various subspace clustering approaches in different ways.

The well known survey specified by P. Lance et al. [13] suggests two major types of subspace clustering as top down and bottom up, based on the search strategy used. The top down approaches are further classified as per cluster weighting methods and per instance weighting methods. However, in this classification, the authors have considered only grid based approaches and have classified them further based on the size of grid as static grid and adaptive grid approaches. No clear division is stated by the authors about bottom up approaches as grid based or density based. We have shown this classification of bottom up approaches in the following sections.

Ilango et al [11] classify high dimensional clustering approaches as partitioning approaches, hierarchical approaches, density based approaches, grid based approaches and model based approaches and further present a survey of various grid based approaches. However, there is no specific categorization about subspace clustering approaches or traditional clustering approaches. Karlton S. et al [33] classify subspace clustering approaches into two categories: density based clustering and projected clustering. As per the authors, density based clustering approaches such as CLIQUE (Clustering In QUEst) [9], MAFIA (Merging Adaptive Finite Intervals And is more than a clique) [28], SUBCLU (density connected SUBspace CLUstering) [14] are based on density of data. Projected clustering is observed in approaches such as PROCLUS (PROjected CLUStering) [30], CLARANS [34], ORCLUS (arbitrarily Oriented projected CLUStering) [35], DOC (Density based Optimal projective Clustering) [31] etc. However, it is not clearly stated by the authors on which basis the density based approaches differ from projected approaches.

H.P. Kriegel et al [10] classify different high dimensional data clustering approaches as subspace clustering (or axis parallel clustering), correlation clustering (arbitrarily oriented clustering) and pattern based clustering. Correlation clustering approaches aim at finding clusters, which may exist in any arbitrarily oriented subspaces, e.g. ORCLUS [35]. Pattern based clustering aims at grouping objects in clusters exhibiting similar trend in a subset of attributes, e.g. p-Cluster [36]. Axis parallel subspace clustering algorithms are further classified using problem oriented categorization as subspace clustering, projected clustering, soft projected clustering and hybrid algorithms. Projected clustering approaches aim at finding a unique assignment of each object to exactly one subspace cluster or noise, e.g. PreDeCon (subspace PREference weighted DEnsity CONnected clustering) [32]. In soft projected clustering algorithms, the number k of clusters is known in advance and an objective function is defined which is optimized to generate k-number of clusters, e.g. COSA (Clustering Objects on Subsets of Attributes) [37]. Subspace clustering algorithms aim at finding all subspaces where clusters can be identified, e.g. SUBCLU [14]; and hybrid algorithms aim at finding something in between, i.e. these algorithms may find overlapping clusters, but may not claim that these will search for every possible subspace and every possible cluster. e.g. FIRES (FIlter REfinement Subspace clustering) [29].

One more simple classification is stated in [38]. Depending on the underlying cluster definition and parameterization of the resulting clusters, authors have classified subspace clustering approaches as cell based, density based and clustering oriented approaches. Cell based approaches search for sets of fixed or variable grid cells containing more than a certain threshold objects, e.g. CLIQUE [9]. Density based approaches define clusters as dense regions separated by sparse regions, e.g. SUBCLU [14] and clustering oriented approaches define properties of the entire set of clusters, like the number of clusters, their average dimensionality or statistically oriented properties, e.g. PROCLUS [30]. In fact, there exist other classifications, which the authors have not taken care of.

## 2.1 Subspace Clustering Algorithm Family (SCAF)

Most of the classification schemes stated in the previous section are not comprehensive in nature. They classify existing approaches into various classes and the classes vary from author to author.

However, a comprehensive classification scheme is necessary which will divide existing clustering approaches into various families, so that the algorithms belonging to one family will satisfy common characteristics. For example - axis parallel, overlapping, bottom up, density based subspace clustering algorithms form one family. This concept will help categorize various existing approaches not only in different classes but also in different families. Thus, a new algorithm proposed by a researcher will be compared only with the algorithms belonging to the respective family.

We propose that, various classes decide the characteristics of a common family so that similar algorithms will belong to that family. With this view, the characteristics of a family will be based on the classes as mentioned in Table 1.

With reference to Table 1, combinations of different classes build different SCAF, such as –

　　i. Axis parallel, overlapping, bottom up, grid based SCAF

　　ii. Axis parallel, overlapping, bottom up, density based SCAF

　　iii. Axis parallel, non-overlapping, bottom up, grid based SCAF

　　iv. Axis parallel, non-overlapping, bottom up, density based SCAF

　　v. Arbitrarily oriented, overlapping, bottom up, grid based SCAF

　　vi. Arbitrarily oriented, overlapping, bottom up, density based SCAF

　　vii. Arbitrarily oriented, non-overlapping, bottom up, grid based SCAF

　　viii. Arbitrarily oriented, non-overlapping, bottom up, density based SCAF

Table 1. SCAF classes and corresponding characteristics

| Sr. No. | Class | Characteristics of the family |
|---|---|---|
| [1] | Cluster orientation | Axis parallel |
|  |  | Arbitrarily Oriented |
| [2] | Overlap of dimensions or objects | Overlapping |
|  |  | Non-overlapping |
| [3] | Search methods | Bottom up |
|  |  | Top down |
| [4] | Use of grid | Grid based |
|  |  | Density based |

Similarly, different combinations of top down approaches could be identified. These classes building various SCAFs are discussed in the following sections.

**2.2 Overlap of Dimensions or Objects - Overlapping / Non-overlapping**

On a very elemental level, subspace clustering algorithms can be classified as overlapping clusters and non-overlapping clusters according to the results they produce. Overlapping clusters allow data points to belong to several clusters in varying subspace projections. Non-overlapping clusters assign each object either to a unique cluster or to a noise.

In Figure 3, Cluster1 and Cluster2 represent overlapping subspace clusters as they share a common object p7. Cluster3, cluster5 are non-overlapping subspace clusters appearing in dimensions {d5, d6, d7} and {d13, d14, d15} respectively. Cluster4 represents a traditional full dimensional cluster.

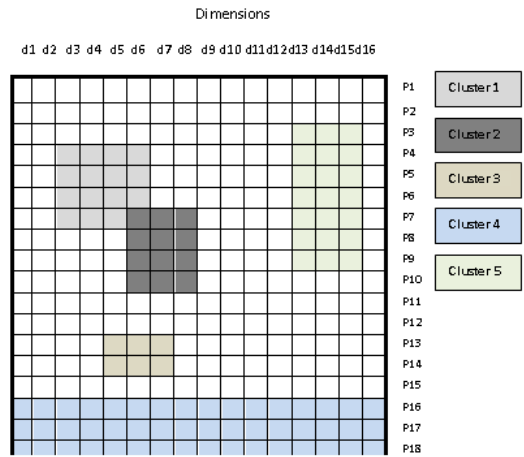

Figure 3. Overlapping and non-overlapping clusters

Overlapping cluster algorithms aim at finding every possible cluster in every possible subspace of the feature space. The major examples of such algorithms include CLIQUE [9], ENCLUS (ENtropy based subspace CLUStering) [27], MAFIA [28], SUBCLU [14], FIRES [29] etc. Significant examples of non-overlapping approaches are PROCLUS [30], DOC [31], PreDeCon [32] etc.

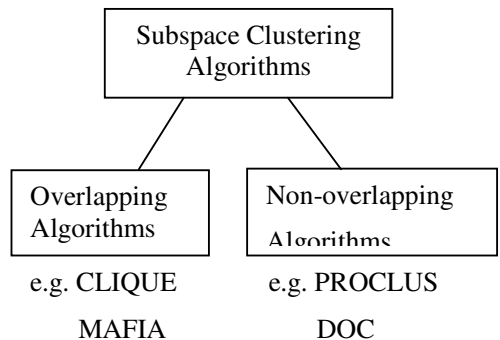

Figure 4. Overlapping / non-overlapping subspace clusters

## 2.3 Search Methods : Bottom Up / Top Down

Another fundamental classification of subspace clustering approaches is based on search techniques employed, such as 'top down' or 'bottom up' [13]. The top down approach starts finding clusters in full feature space by giving equal weights to all dimensions. Each dimension is assigned with a weight for each cluster and iterates multiple times to regenerate the clusters. This continues till it reaches the intended length of subspaces, which is an input parameter. These clusters are partitions of dataset, thus creating non-overlapping clusters, e.g. PROCLUS [30], ORCLUS [35], FINDIT (Fast and INtelligent subspace clustering algorithm using Dimension voTing) [39].

Bottom up search methods are based on downward closure property of density as in apriori property, which reduces the search space significantly [9]. These approaches initially test each dimension for clustering and select those dimensions which have quality above a given threshold. This process subsequently combines these dimensions to build candidate subspaces which can be checked against given threshold value, to prune not so important subspaces. The algorithm stops when it cannot find any more quality subspaces. Few significant algorithms of this category are CLIQUE [9], DOC [31] and MAFIA [28].

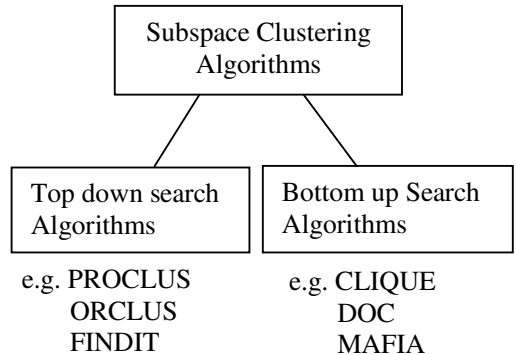

Figure 5. Top down / bottom up search based subspace clustering

Both top down and bottom up approaches are commonly used in the domain of data mining. Both types of algorithms attempt to detect subspace clusters efficiently. Top down approaches first identify cluster members and then decide the related subspaces of these clusters. Whereas, bottom up approaches first predict interesting subspaces and then search for the clusters in those subspaces. Figure 5 presents a classification scheme based on search techniques employed in subspace clustering approaches.

## 2.4 Cluster Orientation – Axis Parallel / Arbitrarily Oriented

Subspace clustering algorithms can also be classified as axis parallel approaches which aim at finding clusters in axis parallel subspaces of the data space, e.g. CLIQUE [9]. Correlation clustering aims at finding clusters in arbitrarily oriented subspaces of the feature space, e.g. ORCLUS [35]. These approaches are also known by the names as correlation clustering, generalized subspace clustering or oriented clustering.

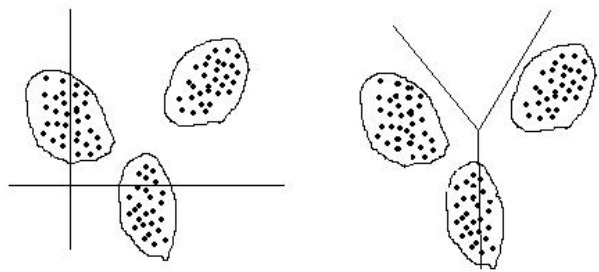

Figure 6. Axis parallel clusters and arbitrarily oriented clusters

Figure 6 shows that, clusters can be better expressed in arbitrarily oriented subspaces. However, if we consider arbitrarily oriented clusters, the computational efficiency goes quite low compared to axis-parallel approaches, as the number of possible subspaces goes to infinite. Thus, depending on the applications, sometimes it is reasonable to simply locate axis parallel clusters, as finding such clusters is more efficient compared to correlation clusters.

## 2.5 Use of Grid – Grid Based / Density Based

Grid based clustering approach is based on cell approximation of data space. This approach uses a multi-resolution grid data structure which quantizes the object space into a finite number of cells forming an axis parallel grid structure. It first partitions the range of values in every dimension into equal sized / variable sized cells and then combines the high density adjacent cells to form a cluster. Resultant clusters thus consist of set of cells containing more than a threshold $\tau$ number of objects.

$$|O_i| \geq \tau \text{ for } i = 1..k, \ k = \text{total number of clusters detected}$$

Figure 3 shows an example of 2-d static grid containing clusters.

CLIQUE [9], the pioneering approach, uses grid based technique for subspace clustering. Few variants of CLIQUE are DOC [31], ENCLUS [27], MAFIA [28], STING (STatistical INformation Grid) [40], etc.

Depending on the grid of fixed width or variable width, these approaches can be further categorized as static or adaptive. Static grid can be regarded as discretization of the data space, whereas adaptive grid can be positioned arbitrarily to maximize the objects in a particular region. MAFIA [28] is the first adaptive grid based clustering approach. AMR (Adaptive Mesh Refinement) [41] is another variant of MAFIA.

In summary, grid based approaches are the first subspace clustering approaches based on simple but efficient cluster model. However, the major limitation of all these techniques is caused by positioning of grid. Shapes of the clusters found, are always a polygon with lines parallel to axes corresponding to subspaces. Sizes of the clusters always depend on orientation of the grid. These algorithms give better efficiency in medium range of dimensions. However, as the number of dimensions increases, the number of cells increases exponentially and this may degrade the efficiency to a large extent. In such cases, density based approaches are found superior.

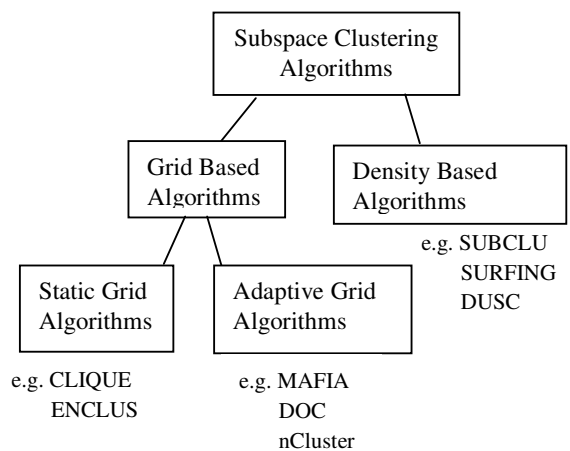

Figure 7. Classification based on use of grid and density notion

Density connected clustering approaches are based on the clustering paradigm specified by DBSCAN [42]. They compute the density around a certain point by searching its $\varepsilon$–neighborhood. It requires two input parameters, $\mu$-threshold and $\varepsilon$–radius, to define whether an area is dense or not. A cluster is then defined as a set of dense objects having more than $\mu$ number of objects in $\varepsilon$–neighborhood. A density based subspace cluster $(O, S)$ with respect to density threshold $\mu$ and $\varepsilon$–radius, can be defined as– an objects $o$ is dense : $\forall o \in DB, |N_\varepsilon(o)| \geq \mu$, where $|N_\varepsilon(o)|$ is the $\varepsilon$–neighborhood of object $o$ and can be derived as –

$$|N_\varepsilon(o)| = \{ \ p \ \epsilon \ DB \ | \ Dist_S(o,p) \leq \varepsilon \ \}.$$

$Dist_S$ designates the distance function applied on a set of dimensions S.

Thus, all core objects together which share common neighbourhood, define the outline of a cluster. Non-core objects within the neighbourhood of core objects form the boundary of the cluster. The objects which do not belong to any of the clusters are regarded as noise points. Density based approaches can thus find clusters of any shape and size and are noise tolerant.
The first density based subspace clustering algorithm is SUBCLU [14]. Other examples are FIRES [29], PreDeCon [32], DUSC (Dimensionality Unbiased Subspace Clustering) [43] and SURFING (SUbspace Relevant For ClusterING) [44].

Density based subspace clustering approaches can detect clusters of any size / shape which may be positioned arbitrarily, thus eliminating the problems associated with grid based approaches. However, as the density measurement is again based on distance, density based clustering approaches compute distances by considering only the relevant dimensions.

Until now, we have discussed individual classes, classifying various subspace clustering algorithms. SCAF helps classify these algorithms into families making it friendlier to researchers.

## 2.6 Advantages of SCAF

In general, comparing and evaluating the results of any subspace clustering algorithm is based on parameters and techniques used. Clustering being an unsupervised learning, there is no information as to which subspaces contains clusters in real life data. As this field of subspace clustering is yet latest and emerging, there is no common ground available to bring all the algorithms on a universal platform, so that their results could be compared. As a result, when a research publication claims that its proposed algorithm offers better results, it might have limited comparison of the results with its preferred algorithms or its preferred paradigm only.

The concept of SCAF will help solve this problem of comparing clustering results against the preferred one. As a future work, we can further identify appropriate comparison measures or evaluation paradigms for each of the family. Whenever, a researcher suggests a new subspace clustering algorithm, he / she will apply various classes and decide the family to which the new algorithm belongs. Each family will have few benchmarking clustering algorithms against which the results of the new algorithm can be compared and tested. Thus the concept of clustering family will present a common basis for research in subspace clustering, enabling the researchers to have appropriate experimentation and testing with correct assumptions.

A comparative study of few significant algorithms belonging to "Axis parallel, overlapping, bottom up, density based" SCAF is presented in the next section. We tested these algorithms using synthetic dataset 'diabetes' consisting of 8 dimensions and 779 instances, which is supported by *OpenSubspace* [38], an open source framework for evaluation and exploration of subspace clustering algorithms in Weka. *OpenSubspace* supports SUBCLU and FIRES which belongs to our example subspace clustering family. So we compared the results of other approaches with FIRES & SUBCLU.

## 3. "AXIS PARALLEL, OVERLAPPING, DENSITY BASED" SCAF

"Axis parallel, overlapping, density based" algorithms are summarized based on their working strategies, heuristics used, data structures used, shape and size of the clusters, run time, accuracy, limitations etc. as follows.

### 3.1 SUBCLU (density-connected SUBspace CLUstering)

SUBCLU [14] is the first density based subspace clustering approach extending the concept of DBSCAN for high dimensional data. It is based on a greedy algorithm to detect the density connected clusters in all subspaces of high dimensional data. It uses monotonicity property to remove higher dimensional projections reducing the search space largely. It overcomes the limitations of grid-based approaches such as dependence on the positioning of the grids or fixed shape of clusters.

The algorithm starts by generating all 1-dimensional clusters using input parameters, μ- density threshold and ε-distance (radius), by applying DBSCAN [42] to each 1-dimensional subspace. Then, it checks for every k-dim cluster, whether it exists in any of (k-1)-dim clusters and if it does not exist, it will be pruned. Lastly, clusters are generated by applying DBSCAN on each (k+1) dimensional candidate subspace. These steps are recursively executed as long as the set of k-dimensional subspaces containing clusters is not empty.

We tested SUBCLU using *OpenSubspace* [38], an open source framework for evaluation and exploration of subspace clustering algorithms in Weka. Compared to FIRES, SUBCLU achieves a better clustering quality; however, it takes longer time to execute.

### 3.2 FIRES (FIlter REfinement Subspace clustering)

FIRES [29] is an efficient subspace clustering algorithm as it uses approximate solution. Rather than going bottom up, it makes use of 1-d histogram information (called base clusters) and jumps directly to interesting subspace regions. Moreover, generation of these base clusters can be done using any clustering approach and may not restrict to DBSCAN.

FIRES then generate cluster approximations by combining base clusters to find maximum dimensional subspace clusters. These clusters are not merged in an apriori style. But FIRES uses an algorithm that scales at most quadratic with respect to the number of dimensions. It refines these cluster approximations as a post processing step to better structure the subspace clusters.

We also tested FIRES using *OpenSubspace* in Weka. Compared to SUBCLU, FIRES takes too small execution time and gives more accurate results than SUBCLU.

### 3.3 DUSC (Dimensionality Unbiased Subspace Clustering)

To overcome the effect of varying dimensionality of subspaces, DUSC [43] gives a formal definition of dimensionality bias, based on statistical foundations.

In density based clustering, density of an object *o* is determined by simply counting the number of objects in a fixed ε-neighbourhood $N_\varepsilon^S(O)$. DUSC generalizes this idea by assigning weights to each object contained in $N_\varepsilon^S(O)$. Thus an object *o* in subapce S is called dense if the weighted distances to objects in its area of influence sum up to more than a given density threshold τ i.e. $\varphi^S(o) \geq \tau$. It further uses Epanechnikov kernel estimator [43] to estimate density value at any position in the data space. It assigns decreasing weights to objects with increasing distance. Density of any object is then measured with respect to the expected density α (S). Thus, an object *o* is dense in subspace S according to the expected density α (S), if and only if,

$$\frac{1}{\alpha(s)}\varphi^S(o) \geq F$$

where F denotes the density threshold. F is independent of the dimensionality and data set size, and is much easier to specify than traditional density thresholds. DUSC also combines the major paradigms for improving the runtime.

The experiments on large high dimensional synthetic and real world data sets show that DUSC outperforms other subspace clustering algorithms in terms of accuracy and runtime.

### 3.4 INSCY (INdexing Subspace Clusters with in-process-removal of redundancY),

INSCY [45] is another efficient subspace clustering algorithm which is based on the subspace clustering notion of [43]. It uses a depth first approach to mine recursively in a region of all clusters in all subspace projections and then continue with the next region. Because of this, it evaluates the maximal high dimensional projection first, quickly pruning all its redundant low dimensional projections. This approach leads to major efficiency gains as it overcomes the drawbacks of breadth first subspace clustering reducing runtimes substantially. Also, this allows indexing of promising subspace cluster regions. INSCY proposes a novel index structure SCY-tree, which provides a compact representation of the data allowing arbitrary access to subspaces. SCY-tree combines in-process redundancy pruning, for very efficient subspace clustering. This makes INSCY fast and concise.

Thorough experiments on real and synthetic data show that INSCY yields substantial efficiency and quality improvements over traditional density based subspace clustering algorithms.

### 3.5 Scalable Density Based Subspace Clustering

Scalable density based subspace clustering [46] is a method that steers mining to few selected subspace clusters only. It reduces subspace processing by identifying and clustering promising subspaces and their combinations directly, narrowing down the search space while maintaining accuracy. It uses the principle that any high dimensional subspace cluster appears in many low dimensional projections. By mining only some of them, the algorithm gathers enough information to jump directly to the more interesting high dimensional subspace clusters without processing the in between subspaces. Database scans are completely avoided with this approach for many intermediate, redundant subspace projections, steering the process of subspace clustering.

It uses priority queue to initialize the information of density estimates. It gives a basis for selecting the best candidate from the priority queue. The priority queue is split into three levels for multiple density granularities. It skips intermediate subspaces in a best first manner and jumps directly to high dimensional subspaces. The experiments prove that the best first selection of subspace clusters enables a scalable subspace clustering algorithm with enhanced run time and it also produces high quality subspace clustering.

### 3.6 DENCOS (DENsity COnscious Subspace clustering)

DENCOS [47] addresses the critical problem of high dimensional data clustering as "density divergence problem" i.e. different subspace cardinalities have different region densities. It uses a novel data structure DFP-tree (Density Frequent Pattern-tree) to save the information of all dense units. It discovers clusters in divide-and-conquer manner using this data structure. DENCOS works in two phases – pre-processing phase where it constructs DFP-tree; and discovering phase where clusters are discovered using dense units' information saved in DFP-tree.

To solve the problem of density divergence, DENCOS formulates a novel subspace clustering model which discovers the clusters based on the relative region densities. Different density thresholds are adaptively calculated to locate the clusters in different subspace dimensionalities. As validated by the extensive experiments on various data sets, DENCOS finds out clusters in all subspaces efficiently.

Section 3 thus covered a reasonable description of all those state of the art algorithms belonging to the same subspace clustering algorithms' family. Further, table 2 summarizes the important aspects, characteristics of all these algorithms.

## 4. CONCLUSION

In this paper, a detailed introduction to cluster analysis in high dimensional data and challenges faced by such clustering approaches are presented. The major challenges in high dimensional data clustering are curse of dimensionality, irrelevant dimensions and correlations among various dimensions. We described in brief, traditional approaches such as feature selection and feature transformation, to solve the problem of high dimensional data clustering. We then presented details about subspace clustering - a most commonly used high dimensional data clustering approach.

Lots of approaches exist for subspace clustering and numerous algorithms are being proposed nearly every day. Proper selection of a clustering approach to suit a particular application and data, should be based on –

   i. Understanding of the exact requirement of clustering application and
   ii. Principles of working of available approaches.

Hence, an attempt is made to present various classification schemes for existing subspace clustering algorithms to better understand group-characteristics of various families of algorithms.

The concept of SCAF, Subspace Clustering Algorithm Family, is presented to help solve the problem of building a uniform platform to classify and hence test new subspace clustering algorithms. Examples of few families are created by assigning different values to classes which define SCAF. A comparative study of few specific algorithms belonging to "Axis parallel, overlapping, density based" SCAF has been presented for ready reference. Their comparison based on different parameters such as run time, shape and size of the clusters etc. is also presented. We implemented few of these techniques on OpenSubspace (Weka) to better understand their working, strengths and limitations; although there is a need for more extensive testing and comparative study of all these techniques.

Finally, it is not possible that every clustering approach will be suitable to every type of data. We limited the scope of this paper only to continuous valued data; though, there exist many clustering algorithms which are specially designed for stream data, graph data, spatial data, text data, heterogeneous data etc. We hope to stimulate further research in these areas.

Table 2. Characteristics of axis parallel, density based subspace clustering algorithms

| Clustering Approach | Overlapping | Bottom-up | Top-down | Input Parameters | Shape of the Cluster | Use of global density threshold | Robust to noise | use of Monotonicity Property | Independent of order of data | Variable length subspaces | Runtime | Special Properties |
|---|---|---|---|---|---|---|---|---|---|---|---|---|
| SUBCLU | Y | Y | - | $\varepsilon$ & $\mu$ | Arb | Y | Y | Y | Y | Y | $O(n \log n)$ | Arbitrary number of clusters |
| FIRES | Y | Y | - | $\varepsilon$ & $\mu$ | Arb | N | Y | Y | Y | Y | $O(n)$ | 1-d histograms to prune dims |
| DUSC | Y | Y | - | $\mu$ | Arb | N | Y | Y | Y | Y | Exponential | Dimensionality unbiased subspace clustering, based on statistical foundation |
| INSCY | Y | Y | - | $\tau$, R & $\varepsilon$ | Arb | N | Y | Y | Y | Y | Exponential | Depth first processing, use of SCY-tree to represent clusters |
| Scalable Density based Subspace Clustering | Y | Y | - | P | Arb | N | Y | Y | Y | Y | $O(2^k|DB|^2)$ k-maximal subspace dimensionlity | Best-first selection of subspace clusters, Scalable to high dimensional data |
| DENCOS | Y | Y | - | $\delta$, $\alpha$ & $k_{max}$ | Arb | N | Y | Y | Y | Y | $O(d^k)$ d – dimensionlity k - cardinality of data set | Use of DFP-tree to save information of dense units, Discovers clusters in divide-and-conquer manner |

$\varepsilon$ – Epsilon radius  
P-jump indicator  
$k_{max}$ - maximal subspace cardinality  
$\mu$, $\tau$ - Density threshold  
$\delta$-equal length intervals  
Arb – Arbitrary shaped clusters  
R-redundancy factor  
$\alpha$-unit strength factor  

**ACKNOWLEDGEMENT**

We would like to thank our student Shweta Daptari, for providing help in implementing and testing the algorithms.

International Journal of Data Mining & Knowledge Management Process (IJDKP) Vol.3, No.2, March 2013
...

**Authors**

Sunita Jahirabadkar is working as Asst. Professor in Computer Department, Cummins College of Engineering, Pune (India) for more than 14 years. She has 10 research publications in various national / international conferences / journals. She is a co-author of the book "e-business" by Oxford Publications. Her areas of interest and research include Data Mining, Artificial Intelligence, Machine Learning, Computer Architectures etc.

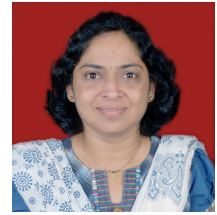

Dr. Parag Kulkarni holds PhD from IIT Kharagpur. UGSM Monarch Business School - Switzerland conferred DSc - Higher Doctorate on him. He is the founder and Chief Scientist of EKLat Research where he has empowered businesses through machine learning, knowledge management, and systemic management. He has been working within the IT industry for over twenty years. The recipient of several awards, Dr. Kulkarni is a pioneer in the field of Systemic Machine Learning. He has over 120 research publications including more than half a dozen books and 3 patents. His areas of research and product development include M-maps, intelligent systems, text mining, image processing, decision systems, forecasting, IT strategy, artificial intelligence, and machine learning.

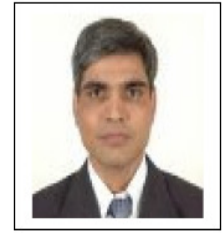